\documentclass[aps,prd,preprint,preprintnumbers]{revtex4}
\usepackage{ae}
\usepackage{bm} 
\usepackage{amssymb}
\usepackage{amsmath}
\usepackage{amsfonts}
\usepackage{slashed}
\usepackage{ulem}
\usepackage{bbold}
\usepackage{graphicx}
\usepackage{hyperref}

\newcommand{\Eqref}[1]{Eq.~\eqref{#1}}

\allowdisplaybreaks

\begin{document}

\date{\today}

\title{Leading low-temperature correction to the Heisenberg-Euler Lagrangian}

\author{Felix Karbstein}\email{felix.karbstein@uni-jena.de}
\affiliation{Helmholtz-Institut Jena, Fr\"obelstieg 3, 07743 Jena, Germany}
\affiliation{GSI Helmholtzzentrum f\"ur Schwerionenforschung, Planckstra\ss e 1, 64291 Darmstadt}
\affiliation{Theoretisch-Physikalisches Institut, Abbe Center of Photonics, \\ Friedrich-Schiller-Universit\"at Jena, Max-Wien-Platz 1, 07743 Jena, Germany}

\begin{abstract}
In this note, we show that the well-known leading low-temperature correction to the Heisenberg-Euler Lagrangian in a constant electromagnetic field arising at two loops can be efficiently extracted from its one-loop zero-temperature analogue.
Resorting to the real-time formalism of equilibrium quantum field theory that explicitly separates out the zero-temperature contribution from the finite-temperature corrections the determination becomes essentially trivial.
In essence, it only requires taking derivatives of the Heisenberg-Euler Lagrangian at one loop and zero temperature for the field strength.
As a bonus, we then effectively dress the low-temperature contribution at two loops by one-particle reducible tadpole structures. This generates a subset of higher-loop contributions to the Heisenberg-Euler Lagrangian in the limit of low temperatures. We extract their leading strong-field behavior at a given loop order, and finally resum these to all loop orders.
\end{abstract}

\maketitle

\section{Introduction}\label{sec:intro}

Quantum electrodynamics (QED) describes the interaction of massive charged Dirac fermions (electrons and positrons of mass $m$ and charge $e$) and massless photons.
By definition, vacuum diagrams do not contain any external particle lines.
This immediately implies that the only physical scale inherited by vacuum diagrams in QED is $m$, and that these can be organized in a formal power series in the fine structure constant $\alpha=e^2/(4\pi)$ at the energy scale of the electron mass $m$, i.e., $\alpha\equiv\alpha(m^2)\simeq1/137$.
Throughout this work, we use the Heaviside-Lorentz System with natural units $c=\hbar=k_{\rm B}=1$ and the metric convention $g^{\mu\nu}={\rm diag}(-1,+1,+1,+1)$; see \cite{Karbstein:2023} for the detailed conventions.

In the presence of a constant external electromagnetic field $F^{\mu\nu}={\rm const}.$, specified in terms of the field strength tensor $F^{\mu\nu}=\partial^\mu A^\nu(x)-\partial^\nu A^\mu(x)$ with associated gauge potential $A^\mu(x)$, the charged particle lines in the vacuum diagrams can couple arbitrary even powers of $eF^{\mu\nu}$ \cite{Heisenberg:1935qt,Weisskopf:1996bu,Schwinger:1951nm}.
This effectively supplements the theory governing the external electromagnetic field, that would be linear in the classical vacuum, with nonlinearities which can be attributed to the quantum vacuum.
Such studies were pioneered by Heisenberg and Euler \cite{Heisenberg:1935qt} who explicitly evaluated the leading loop correction $\sim\alpha^0$ to classical Maxwell theory at zero temperature ($T=0$) to all orders in $eF^{\mu\nu}$.
They presented their result in terms of an Lagrangian that is now known as one-loop Heisenberg-Euler effective Lagrangian ${\cal L}_{\rm HE}^{1\text{-loop}}$ \cite{Dunne:2004nc}.
Nowadays also the complete expression for ${\cal L}_{\rm HE}^{2\text{-loop}}$ at $T=0$ is known \cite{Ritus:1975pcc,Gies:2016yaa}.
As a caveat, we note that beyond one loop there are vacuum diagrams that vanish for structural reasons in the absence of an external field, but contribute in its presence \cite{Gies:2016yaa}.

Reference~\cite{Gies:1999vb} emphasized that, on first sight surprisingly, for temperatures $T\ll m$ the leading finite-temperature corrections to the Heisenberg-Euler effective Lagrangian ${\cal L}_{\rm HE}$ is not encoded in ${\cal L}_{\rm HE}^{1\text{-loop}}$ but stems from ${\cal L}_{\rm HE}^{2\text{-loop}}$, implying a {\it two-loop dominance}; see also \cite{Dittrich:2000zu}.
Employing the imaginary-time formalism of equilibrium quantum field theory (QFT), it in particular determined the finite-temperature corrections $\sim T^4$ and $\sim T^6$.
Using properties of the real-time formalism, subsequently we briefly recall the basic argument why this is the case, and then repeat this exercise in Sec.~\ref{sec:results1}. In Sec.~\ref{sec:results2} we extend these considerations to a certain class of higher-loop diagrams and finally conclude in Sec.~\ref{sec:concls}.

In the real-time formalism of equilibrium QFT the tree-level propagators at finite temperature $T$ naturally decompose into the well-known zero-temperature result and an additional contribution accounting for the finite-temperature corrections \cite{Dolan:1973qd,Donoghue:1983qx}.
In momentum space the latter is proportional to $\delta\bigl(k^2+M^2\bigr)/\bigl({\rm e}^{\beta|k^0|}\mp1\bigr)$ with four-momentum $k^\mu=(k^0,\vec{k})$, particle mass $M$ and inverse temperature $\beta=1/T$.
Here, the $-$ sign holds for bosons and the $+$ sign for fermions, as these are governed by Bose-Einstein and Fermi-Dirac statistics, respectively.
From the above scaling it is clear that for massive particles and sufficiently low temperatures $T\ll M$ the finite-temperature corrections are exponentially suppressed by a factor of ${\rm e}^{-M/T}\ll1$ relative to the zero-temperature contribution.
For massless particles, such as photons, the situation is different as soft momentum modes fulfilling $|\vec{k}|\lesssim T$ remain unsuppressed.

\begin{figure}
 \centering
 \includegraphics[width=0.65\textwidth]{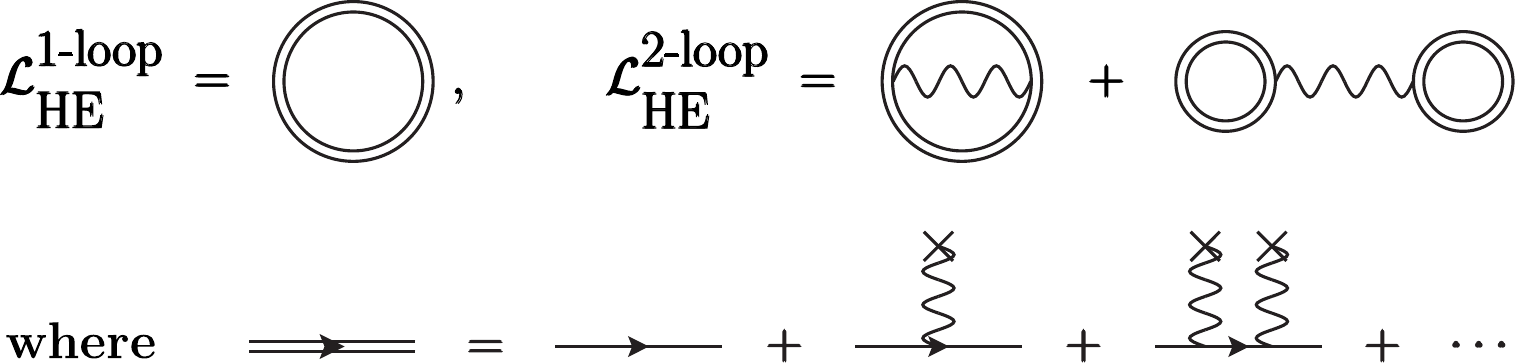}\vspace*{-1mm}
 \caption{Diagrammatic representation of the Heisenberg-Euler effective Lagrangian at one and two loops. The double solid line denotes the Dirac propagator dressed to all orders in $eF^{\mu\nu}$ (wiggly lines ending at crosses). The wiggly lines in the diagrams in the first row represent tree-level photon propagators~\eqref{eq:D}.}
 \label{fig:1}
\end{figure}

In the context of QED vacuum diagrams, this suggests that in the limit of $T\ll m$ the leading finite-temperature corrections should arise from diagrams with photon lines.
Clearly, vacuum diagrams that couple external electromagnetic fields and contain photon lines scale at least linearly with $\alpha$. Because of ${\cal L}_{\rm HE}^{\ell\text{-loop}}\sim\alpha^{\ell-1}$ the lowest-order finite-temperature corrections for $T\ll m$ should thus stem from ${\cal L}^{2\text{-loop}}_{\rm HE}$, which can be expressed as \cite{Dittrich:1985yb,Gies:2016yaa,Karbstein:2023}
\begin{multline}
 {\cal L}^{2\text{-loop}}_{\rm HE}=\frac{\rm i}{V^{(4)}}\,\frac{1}{2}\int\frac{{\rm d}^4k}{(2\pi)^4}\int\frac{{\rm d}^4k'}{(2\pi)^4}\,\Pi^{1\text{-loop}}_{\mu\nu}(k,k')\,D^{\mu\nu}(k,k') \\
 +2\int\frac{{\rm d}^4k}{(2\pi)^4}\int\frac{{\rm d}^4k'}{(2\pi)^4}\,(2\pi)^4\delta(k)\,k^\rho\frac{\partial{\cal L}_{\rm HE}^{1\text{-loop}}}{\partial F^{\rho\mu}} D^{\mu\nu}(k,k')  \frac{\partial{\cal L}_{\rm HE}^{1\text{-loop}}}{\partial F^{\nu\sigma}} k'^\sigma
 \label{eq:L2loops}
\end{multline}
in terms of the one-loop photon polarization tensor $\Pi^{1\text{-loop}}_{\mu\nu}$ evaluated in the background field $F^{\mu\nu}={\rm const}.$, derivatives of ${\cal L}_{\rm HE}^{1\text{-loop}}$ and the photon propagator $D^{\mu\nu}$ in momentum space; $V^{(4)}$ denotes space-time volume in $d=3+1$ dimensions. See Fig.~\ref{fig:1} for a graphical representation of ${\cal L}_{\rm HE}^{1\text{-loop}}$ and the two contributions to ${\cal L}_{\rm HE}^{2\text{-loop}}$ in \Eqref{eq:L2loops}. The contribution in the first line of \Eqref{eq:L2loops} is one-particle irreducible (1PI), and the one in its second line is one-particle reducible (1PR): the corresponding Feynman diagram can be split into two by cutting a single line.
We emphasize that because of $T/m=1.69\times10^{-7}(T/{1{\rm K}})$, the criterion $T\ll m$ is fulfilled for a very wide range of temperatures.
At the same time, the parametric suppression with powers of  $T/m\ll1$ typically renders finite-temperature corrections to the Heisenberg-Euler Lagrangian extremely small and therefore essentially inaccessible in experiment.
However, for future precision studies of the properties of light emitted from magnetars producing extremely strong magnetic fields and featuring surface temperatures of the order of $10^6\,{\rm K}$ \cite{Capparelli:2017mlv} such finite-temperature corrections may ultimately become relevant.

In Feynman gauge, the tree-level photon propagator in momentum space at finite temperature can be expressed as \cite{Dolan:1973qd,Donoghue:1983qx}
\begin{equation}
 D^{\mu\nu}(k,k')=(2\pi)^4\delta(k+k')\,g^{\mu\nu}\biggl(\frac{1}{k^2-{\rm i}0^+}+2\pi\delta(k^2)\frac{\rm i}{{\rm e}^{\beta|\vec{k}|}-1}\biggr).
 \label{eq:D}
\end{equation}
Note that from this expression it is immediately obvious that the contribution in the second line of \Eqref{eq:L2loops} corresponding to the 1PR diagram in Fig.~\ref{fig:1} will not receive a nonzero finite-temperature correction from the thermalized photon line. In fact, at zero temperature the 1PR diagram does not vanish precisely because the infrared (IR) divergence $\sim1/k^2$ of the photon propagator compensates the factor quadratic in $k^\mu$ multiplying the delta function $\delta(k)$ in the numerator \cite{Gies:2016yaa}. Conversely, the finite temperature contribution in \Eqref{eq:D} is manifestly IR finite.
We emphasize that this argument holds for all photon lines connecting tadpole structures, i.e. any substructures that can be split from a given vacuum Feynman diagram by cutting a single photon line.

\section{Results}\

To simplify notations, from now on we explicitly split the full constant field effective Lagrangian as ${\cal L}_{\rm HE}\,\to\,{\cal L}_{\rm HE}+{\cal L}_{\rm HE}^T$ into its zero-temperature limit ${\cal L}_{\rm HE}$ and an associated purely thermal part ${\cal L}_{\rm HE}^T$.
The same splitting ${\cal L}_{\rm HE}^{\ell\text{-loop}}\,\to\,{\cal L}_{\rm HE}^{\ell\text{-loop}}+{\cal L}_{\rm HE}^{\ell\text{-loop},T}$ is adopted at a given loop order $\ell$, and for the photon propagator $D^{\mu\nu}\,\to\,D^{\mu\nu}+D^{\mu\nu,T}$.
The entire field dependence of ${\cal L}_{\rm HE}$ and ${\cal L}_{\rm HE}^{\ell\text{-loop}}$ can then be encoded in the gauge-invariant Lorentz scalars ${\cal F}=F_{\mu\nu}F^{\mu\nu}/4=(\vec{B}^2-\vec{E}^2)/2$ and ${\cal G}=F_{\mu\nu}{}^\star\!F^{\mu\nu}/4=-\vec{B}\cdot\vec{E}$ with dual field strength tensor ${}^\star\! F_{\mu\nu}=\epsilon_{\mu\nu\rho\sigma}F^{\rho\sigma}/2$ \cite{Heisenberg:1935qt,Weisskopf:1996bu,Schwinger:1951nm}.

\subsection{Low-temperature correction at two loops}\label{sec:results1}

In the conventions just introduced, the leading finite temperature correction to the Heisenberg-Euler effective Lagrangian for $T\ll m$ is encoded in
\begin{equation}
 {\cal L}^{2\text{-loop},T}_{\rm HE}=\frac{\rm i}{V^{(4)}}\,\frac{1}{2}\int\frac{{\rm d}^4k}{(2\pi)^4}\int\frac{{\rm d}^4k'}{(2\pi)^4}\,\Pi^{1\text{-loop}}_{\mu\nu}(k,k')\,D^{\mu\nu,T}(k,k')\,+\,{\cal O}\bigl({\rm e}^{-m/T}\bigr)\,,
 \label{eq:L2loopsT}
\end{equation}
which we aim to evaluate in the low-temperature limit.
To this end, we first recall that the low-energy limit of the one-loop photon polarization tensor in a constant electromagnetic field can be readily determined from the one-loop Lagrangian \cite{Heisenberg:1935qt,Schwinger:1951nm}
\begin{equation}
 {\cal L}_{\rm HE}^{1\text{-loop}}=-\frac{1}{8\pi^2}\int_{{\rm i}0^+}^{\infty+{\rm i}0^+}\frac{{\rm d}s}{s}\,{\rm e}^{-m^2s}\biggl(\frac{(ec_+)(ec_-)}{\tanh(ec_+s)\tan(ec_-s)}-\frac{1}{s^2}-\frac{(ec_+)^2-(ec_-)^2}{3}\biggr)
 \label{eq:L_HE_1loop}
\end{equation}
by functional differentiation \cite{Karbstein:2015cpa}. Here, $c_\pm=\bigl(\sqrt{{\cal F}^2+{\cal G}^2}\pm{\cal F}\bigr)^{1/2}$. Its momentum space representation follows as $\Pi^{1\text{-loop}}_{\mu\nu}(k,k')=-\frac{\delta}{\delta A^\mu(k)}\frac{\delta}{\delta A^\nu(k')}\int{\rm d}^4x\,{\cal L}_{\rm HE}^{1\text{-loop}}$ with $A^\mu(k)$ denoting the Fourier transform of $A^\mu(x)$, such that
\begin{multline}
 \Pi^{1\text{-loop}}_{\mu\nu}(k,k')
 =-(2\pi)^4\delta(k+k') \biggl[
 \bigl(k^2g_{\mu\nu} - k_\mu k_\nu \bigr)\frac{\partial}{\partial{\cal F}}
 + (kF)_\mu  (kF)_\nu\,\frac{\partial^2}{\partial{\cal F}^2} \\
 + (k{}^\star\!F)_\mu (k{}^\star\!F)_\nu\,\frac{\partial^2}{\partial{\cal G}^2}
 + \bigl[(k{}^\star\!F)_\mu (kF)_\nu + (k F)_\mu (k{}^\star\!F)_\nu\bigr]\,\frac{\partial^2}{\partial{\cal F}\partial{\cal G}}
 \biggr]{\cal L}_{\rm HE}^{1\text{-loop}}\,.
 \label{eq:Pi}
\end{multline}
Equation~\eqref{eq:Pi} provides the exact result for the photon polarization tensor at quadratic order in $k^\mu$ but inherently neglects corrections scaling at least quartically with $k^\mu$.
Its trace entering \Eqref{eq:L2loopsT} is given by
\begin{multline}
 g^{\mu\nu}\Pi^{1\text{-loop}}_{\mu\nu}(k,k')
 =-(2\pi)^4\delta(k+k') \biggl[k^2
 \Bigl(3\frac{\partial}{\partial{\cal F}} + 2{\cal G}\,\frac{\partial^2}{\partial{\cal F}\partial{\cal G}}-2{\cal F}\,\frac{\partial^2}{\partial{\cal G}^2}\Bigr) \\
 + (kF)^2 \Bigl(\frac{\partial^2}{\partial{\cal F}^2} + \frac{\partial^2}{\partial{\cal G}^2}\Bigr)
 \biggr]{\cal L}_{\rm HE}^{1\text{-loop}}\,,
 \label{eq:TracePi}
\end{multline}
where $(kF)^2=(kF)^\rho(kF)_\rho$. In particular, note that $(kF)^2|_{k^0=\pm|\vec{k}|}=(\vec{k}\times\vec{B})^2+(\vec{k}\times\vec{E})^2\pm2|\vec{k}|\vec{k}\cdot(\vec{B}\times\vec{E})$.
Another important observation is the identity
\begin{align}
 \int_0^\infty{\rm dk}\,{\rm k}^{2n}\frac{\rm k}{{\rm e}^{\beta{\rm k}}-1}&=T^{2+2n}\,(2n+1)!\,\zeta(2+2n)\quad\text{for}\quad n\in\mathbb{N}_0\,, \nonumber\\
&=T^{2+2n}\,(-1)^n\,\frac{(2\pi)^{2+2n}}{4(1+n)}\,{\cal B}_{2+2n}\,,
 \label{eq:radints}
\end{align}
where $\zeta(n)$ is the Riemann zeta function and ${\cal B}_n$ are Bernoulli numbers.

Upon insertion of the thermal part of \Eqref{eq:D} and \Eqref{eq:TracePi} into \Eqref{eq:L2loopsT} we obtain
\begin{align}
 {\cal L}^{2\text{-loop},T}_{\rm HE}&=\frac{1}{2}\int\frac{{\rm d}^3k}{(2\pi)^3} \biggl[\Bigl(\frac{\partial^2}{\partial{\cal F}^2} + \frac{\partial^2}{\partial{\cal G}^2}\Bigr){\cal L}^{1\text{-loop}}_{\rm HE}+{\cal O}\bigl(|\vec{k}|^2\bigr)\biggr]\frac{|\vec{k}|}{{\rm e}^{\beta|\vec{k}|}-1} [(\hat{\vec{k}}\times\vec{B})^2+(\hat{\vec{k}}\times\vec{E})^2] 
  \label{eq:L2loopsTtothe4tbc}\\
 \intertext{with unit momentum vector $\hat{\vec{k}}=\vec{k}/|\vec{k}|$. Using spherical coordinates in momentum space and performing the integration over $|\vec{k}|\to{\rm k}$ with \Eqref{eq:radints}  we finally arrive at}
 &=\frac{\pi^2}{45}\,T^4\,{\cal U}\biggl[\Bigl(\frac{\partial^2}{\partial{\cal F}^2} + \frac{\partial^2}{\partial{\cal G}^2}\Bigr){\cal L}^{1\text{-loop}}_{\rm HE}+{\cal O}(T^2)\biggr]\,,
 \label{eq:L2loopsTtothe4}
\end{align}
where we introduced the energy density of the electromagnetic field in the heat-bath rest frame ${\cal U}=(\vec{B}^2+\vec{E}^2)/2$.
Given the structure of \Eqref{eq:L_HE_1loop} depending on $\cal F$ and $\cal G$  only via $c_+$ and $c_-$, the sum of the second derivatives for $\cal F$ and $\cal G$ in \Eqref{eq:L2loopsTtothe4} can be alternatively represented as 
\begin{equation}
 \frac{\partial^2}{\partial{\cal F}^2}+\frac{\partial^2}{\partial{\cal G}^2}=\frac{1}{c_+^2+c_-^2}\Bigl(\frac{\partial^2}{\partial c_+^2}+\frac{\partial^2}{\partial c_-^2}\Bigr)\,.
 \label{eq:FGtocpm}
\end{equation} 
With this substitution \Eqref{eq:L2loopsTtothe4} matches the expression given in Eq.~(30) of \cite{Gies:1999vb}.
For completeness, we also note that in an external electromagnetic field characterized by a single direction, such as (anti-)parallel magnetic and electric fields, or purely magnetic ($c_+=|\vec{B}|$, $c_-=0$)  or electric fields ($c_+=0$, $c_-=|\vec{E}|$), we have
$c_+^2+c_-^2=2\,{\cal U}$.

Higher-order contributions in $T$ can be systematically determined by utilizing the photon polarization tensor in generic constant electromagnetic fields beyond the low-energy limit \cite{Batalin:1971au} in \Eqref{eq:L2loopsT}: As obvious from Eqs.~\eqref{eq:radints}, \eqref{eq:L2loopsTtothe4tbc} and \eqref{eq:L2loopsTtothe4}, the contribution to the photon polarization tensor in momentum space scaling with the $2n$th power of the transferred momentum $k^\mu$ encodes the finite-temperature correction $\sim T^{2+2n}$. We have explicitly confirmed that in this way we indeed recover the expression for ${\cal L}^{2\text{-loop},T}_{\rm HE}$ at ${\cal O}(T^6)$ of \cite{Gies:1999vb}.

Plugging the all-order weak-field expansion of ${\cal L}_{\rm HE}^{1\text{-loop}}$  \cite{Dunne:2004nc,Karbstein:2019oej} into \Eqref{eq:L2loopsTtothe4} and using \Eqref{eq:FGtocpm}, we readily arrive at the following expression,
\begin{align}
 {\cal L}_{\rm HE}^{\text{2-loop},T}&\sim \frac{\pi^2}{45}T^4\frac{\alpha}{\pi}\frac{2\,{\cal U}}{c_+^2+c_-^2}\sum_{n=1}^\infty\sum_{k=0}^n(-1)^k\frac{(2n-1)!}{(2k)!(2n-2k)!} \nonumber\\
 &\quad\quad\times\bigl[{\cal B}_{2(k+1)}{\cal B}_{2(n-k)}-{\cal B}_{2k}{\cal B}_{2(n+1-k)}\bigr]\Bigl(\frac{2ec_+}{m^2}\Bigr)^{2(n-k)}\Bigl(\frac{2ec_-}{m^2}\Bigr)^{2k}+{\cal O}(T^6)\,. \label{eq:pertweakfields}
\end{align}
Note that the entire field dependence of the lowest-order contribution to \Eqref{eq:pertweakfields} is encoded in $\cal U$; the $n=1$ term of the double sum yields a factor of $c_+^2+c_-^2$ canceling out its analog in the denominator.
On the other hand, upon inserting the exact result for the imaginary part of ${\cal L}_{\rm HE}^{1\text{-loop}}$ \cite{Nikishov:1969tt,Karbstein:2019oej} into \Eqref{eq:L2loopsTtothe4}, we find
\begin{align}
 {\rm Im}\bigl\{{\cal L}_{\rm HE}^{\text{2-loop},T}\bigr\}&= \frac{\pi^2}{90}T^4\alpha\,\frac{2\,{\cal U}}{c_+^2+c_-^2}\sum_{n=1}^\infty{\rm e}^{-n\pi\frac{m^2}{ec_-}}\,\frac{\frac{c_+}{c_-}n\pi}{\sinh(\frac{c_+}{c_-}n\pi)} \nonumber\\
 &\quad\quad\times\biggl\{\frac{m^2}{ec_-}\biggl[\frac{1}{2}\frac{m^2}{ec_-}\cosh(\tfrac{c_+}{c_-}n\pi)+\frac{\frac{c_+}{c_-}}{\sinh(\frac{c_+}{c_-}n\pi)}\biggr] \nonumber\\
 &\quad\quad\quad\quad\,\ +\biggl[\frac{c_+^2+c_-^2}{c_-^2}\frac{1}{\tanh(\frac{c_+}{c_-}n\pi)}-\frac{1}{\frac{c_+}{c_-}n\pi}\biggr]\frac{1}{\sinh(\frac{c_+}{c_-}n\pi)}\biggr\}+{\cal O}(T^6)\,. \label{eq:ImL2}
\end{align}
For the purely electric field case ($c_+=0$, $c_-=|\vec{E}|$) this expression simplifies to
\begin{equation}
 \to\, \frac{\pi^2}{90}T^4\alpha\sum_{n=1}^\infty{\rm e}^{-n\pi\frac{m^2}{e|\vec{E}|}}\biggl[\frac{m^2}{ec_-}\Bigl(\frac{1}{2}\frac{m^2}{ec_-}+\frac{1}{n\pi}\Bigr)+\frac{1}{3}+\frac{1}{(n\pi)^2}\biggr]+{\cal O}(T^6)\,,
\end{equation}
which matches the expression obtained in Eq.~(63) of \cite{Gies:1999vb}.
In passing, we note that the explicit determination of the leading finite-temperature correction to ${\rm Im}\{{\cal L}_{\rm HE}^{\text{2-loop},T}\}$ in the special cases of either a {\it magnetic-like field} (i.e., a field for which $c_+\neq 0$ while $c_-=0$) or a {\it constant crossed field} (i.e., orthogonal electric and magnetic fields of the same strength for which $c_+=c_-=0$) is clearly beyond the scope of our present considerations: the contribution given in \Eqref{eq:ImL2} vanishes in both cases.
Reference~\cite{King:2012kd} found the leading finite temperature correction in a constant crossed field to be additionally suppressed by an exponential factor that is non-perturbative in $T/m$. A similar behavior is to be expected for the case of a magnetic-like field.

Moreover, we note that in the special case where ${\cal G}=0$, which encompasses purely electric and magnetic fields, the derivatives entering \Eqref{eq:L2loopsTtothe4} have the following closed form representations \cite{Karbstein:2015cpa}
\begin{align}
 \frac{\partial^2{\cal L}_{\rm HE}^\text{1-loop}}{\partial{\cal F}^2}
 &=\frac{1}{{\cal F}}\frac{\alpha}{\pi}\biggl\{\frac{1}{6}-\biggl[\zeta'(0,\chi)+\frac{1}{2}\ln\chi+\bigl(1-\psi(\chi)\bigr)\chi-\frac{1}{2}\biggr]\chi\biggr\}\,,
 \nonumber\\
 \frac{\partial^2{\cal L}_{\rm HE}^\text{1-loop}}{\partial{\cal G}^2}
 &= \frac{1}{{\cal F}}\frac{\alpha}{\pi}\biggl\{2\zeta'(-1,\chi)-\biggl[\zeta'(0,\chi)-\frac{1}{2}\ln\chi+\frac{\chi}{2}\biggr]\chi -\frac{1}{6}\biggl[\psi(\chi)+\frac{1}{2\chi}+\frac{1}{2}\biggr]\biggr\}\,. \label{eq:d2F+d2G}
\end{align}
Here, we introduced the shorthand notation $\chi=m^2/(2e\sqrt{2{\cal F}})$,  with the square root to be interpreted as $\sqrt{\cal F}=\sqrt{|{\cal F}|}\bigl[\Theta({\cal F})-{\rm i}\Theta(-{\cal F})\bigr]$.
Upon addition, these expressions constitute the closed form representation of \Eqref{eq:L2loopsTtothe4} in the considered limit; see also Eq.~(35) of \cite{Gies:1999vb}.
Utilizing the series expansions of the derivatives $\zeta'(n,\chi)=\partial\zeta(n,\chi)/\partial n$ of the Hurwitz zeta function $\zeta(n,\chi)$ and the digamma function $\psi(\chi)$ for small arguments, these expressions allow to straightforwardly infer the following strong field expansions
\begin{align}
 \frac{\partial^2{\cal L}_{\rm HE}^\text{1-loop}}{\partial{\cal F}^2}
 &=\frac{1}{{\cal F}}\frac{\alpha}{\pi}\biggl\{\frac{1}{6}+\frac{1}{2}\bigl[\ln\chi+\ln(2\pi)-1\bigr]\chi-\chi^2+\sum_{j=0}^\infty(-1)^j\,\frac{j+1}{j+2}\,\zeta(j+2)\chi^{j+3}\biggr\}\,,
 \nonumber\\
 \frac{\partial^2{\cal L}_{\rm HE}^\text{1-loop}}{\partial{\cal G}^2}
 &= \frac{1}{{\cal F}}\frac{\alpha}{\pi}\biggl\{\frac{1}{12\chi}+\frac{\gamma}{6}+2\zeta'(-1)-\frac{1}{12}  -\frac{1}{2}\biggl[\ln\chi+\ln(2\pi)-2+\frac{\pi^2}{18}\biggr]\chi \nonumber\\
 &\hspace*{1.8cm}+\biggl[\frac{1}{2}+\frac{\zeta(3)}{6}\biggr]\chi^2 -\sum_{j=0}^\infty(-1)^j \biggl[\frac{(j+1)\zeta(j+2)}{(j+2)(j+3)}+\frac{\zeta(j+4)}{6}\biggr]\chi^{j+3}\biggr\}\,. \label{d2F+d2Gsfexp}
\end{align}
Interestingly, upon addition of these two expressions the logarithmic terms drop out,  such that -- aside from the leading contribution for $\chi\ll1$ scaling as $1/\chi$ -- when only one of the invariants $c_\pm$ remains finite the strong field expansion of \Eqref{eq:L2loopsTtothe4} can be expressed as a power series in $\chi$.
In the limit of $\chi\ll1$ we find
\begin{equation}
 {\cal L}_{\rm HE}^{\text{2-loop},T}=\frac{\pi}{270}\,T^4\alpha\,\frac{\cal U}{\cal F}\biggl(\frac{e\sqrt{2{\cal F}}}{m^2}+{\cal O}(\chi^0)+{\cal O}(T^2)\biggr)\,, \label{eq:L2loopTsflimit}
\end{equation}
which is in line with Eq.~(36) of \cite{Gies:1999vb}.

\subsection{A glimpse on low-temperature corrections from higher-loop 1PR diagrams}\label{sec:results2}

Subsequently, we stick to the strong field limit and study low-temperature corrections to ${\cal L}_{\rm HE}^T$ that scale quartic in $T\ll m$ and are entirely due to the 1PR sector of the theory. The 1PR (1PI) sector of the theory is formed by the set of all 1PR (1PI) Feynman diagrams with any number of loops contributing to the Heisenberg-Euler Lagrangian.
To be specific, here we limit ourselves to the study of the class of 1PR diagrams containing ${\cal L}_{\rm HE}^{2\text{-loop},T}$ as a building block.
In other words, this amounts to the set of all 1PR diagrams generated by effectively dressing ${\cal L}_{\rm HE}^{2\text{-loop},T}$ with generic tadpole structures.
After identifying the dominant diagram in this subset  at a given loop order we resum these to all orders in the perturbative loop expansion.
From the outset, we emphasize that -- as opposed to the zero-temperature limit where the 1PR sector of the theory can be shown to dominate over the 1PI one \cite{Karbstein:2019wmj} -- we are not aware of any indication that the 1PR sector to be studied here should dominate over its 1PI counterpart. In fact, the presence of $\ell$-loop 1PI diagrams scaling favorably with the field strength for  $\ell>2$ with respect to the 1PR diagrams generated by the dressing of ${\cal L}_{\rm HE}^{2\text{-loop},T}$ with tadpole structures would immediately imply that the latter set of diagrams does not even dominate the 1PR sector of the theory in a strong field. 

Our motivation for the present considerations is twofold: first, we are convinced that -- aside from being much easier to analyze and tackle than the 1PI sector of the theory, which will be studied in a separate work -- the 1PR sector  is of interest in its own right.
Second, already this allows us to explicitly demonstrate that the finite-temperature correction $\sim T^4$ to the Heisenberg-Euler Lagrangian receives contributions $\sim\alpha^{\ell-1}$ from all loop orders $\ell\geq2$.

For simplicity, in the remainder of this work, we focus on the specific situation where not only the condition ${\cal G}=0$ holds, but where the electromagnetic field is either purely magnetic or purely electric in the heat-bath rest frame, i.e., $\sqrt{2{\cal F}}$ may be either $|\vec{B}|$ or $-{\rm i}|\vec{E}|$. This implies $|{\cal U}|=|{\cal F}|$ and comes with additional simplifications because then, as for ${\cal L}_{\rm HE}$ at $T=0$, the field dependence of $ {\cal L}_{\rm HE}^{\text{2-loop},T}$ is effectively encoded in $\cal F$ only. In this context also note the identity $F^{\mu\nu}\frac{\partial}{\partial F^{\mu\nu}}\bigl(\frac{(kF)^2}{\cal F}\bigr)=0$.

Equations~\eqref{eq:L2loopsTtothe4} and \eqref{d2F+d2Gsfexp} predict that at ${\cal O}(T^4)$ the $n$th derivative of ${\cal L}_{\rm HE}^{2\text{-loop},T}$ for $\sqrt{2{\cal F}}$ in the strong magnetic/electric field limit characterized by $|\chi|\ll1\ \leftrightarrow\ |{\cal F}|\gg (m^2/e)^2$ scales as
\begin{equation}
 \frac{\partial^n{\cal L}_\text{HE}^{2\text{-loop},T}}{\partial(\sqrt{2{\cal F}})^n}\sim\alpha^{n/2}\bigl(e\sqrt{2{\cal F}}\bigr)^{-1-n}
 \begin{cases}
  \bigl(e\sqrt{2{\cal F}}\bigr)^2 &\text{for}\quad 0\leq n\leq1 \\
  \quad\ \ 1&\text{for}\quad n\geq2\
 \end{cases}
 \label{eq:dnLT/dBn}
\end{equation}
with respect to the coupling $e\sim\sqrt{\alpha}$ and the renormalization group invariant $e\sqrt{2{\cal F}}$. This behavior of the leading finite-temperature correction to the Heisenberg-Euler Lagrangian at lowest order in both a low-temperature $T\ll m$ and a perturbative loop expansion is to be compared with the behavior of its zero-temperature counterpart ${\cal L}_{\rm HE}^{1\text{-loop}}$.
For $|{\cal F}|\gg(m^2/e)^2$ and ${\cal G}=0$ the derivatives of the latter scale as \cite{Dittrich:1985yb,Dunne:2004nc}
\begin{equation}
 \frac{\partial^n{\cal L}_\text{HE}^{1\text{-loop}}}{\partial(\sqrt{2{\cal F}})^n}\sim\alpha^{n/2}\bigl(e\sqrt{2{\cal F}}\bigr)^{2-n}
 \begin{cases}
  \ln\bigl(\tfrac{e\sqrt{2{\cal F}}}{m^2}\bigr) &\text{for}\quad 0\leq n\leq2 \\
  \quad\ \ 1&\text{for}\quad n\geq3\
 \end{cases}\,.
 \label{eq:dnL/dBn}
\end{equation}
The derivatives in Eqs.~\eqref{eq:dnLT/dBn} and \eqref{eq:dnL/dBn} determine the leading 1PI effective couplings $\sim T^0$ and $\sim T^4$ of $n$ tadpole structures in the strong magnetic/electric field limit \cite{Karbstein:2021gdi}.
In turn, Eqs.~\eqref{eq:dnLT/dBn} and \eqref{eq:dnL/dBn} allow to assess the relative importance of different 1PR diagrams to the Heisenberg-Euler Lagrangian in the strong field limit.
A comparison of Eqs.~\eqref{eq:dnLT/dBn} and \eqref{eq:dnL/dBn} then unveils the following behavior,
\begin{equation}
 \frac{\partial^n{\cal L}_\text{HE}^{2\text{-loop},T}}{\partial(\sqrt{2{\cal F}})^n}\biggl(\frac{\partial^n{\cal L}_\text{HE}^{1\text{-loop}}}{\partial(\sqrt{2{\cal F}})^n}\biggr)^{-1}\sim\bigl(e\sqrt{2{\cal F}}\bigr)^{-3}
 \begin{cases}
  \bigl(e\sqrt{2{\cal F}}\bigr)^{2}\ln^{-1}\bigl(\tfrac{e\sqrt{2{\cal F}}}{m^2}\bigr) \ &\text{for}\quad 0\leq n\leq1 \\
  \quad\quad\ln^{-1}\bigl(\tfrac{e\sqrt{2{\cal F}}}{m^2}\bigr) &\text{for}\quad n=2\ \\
  \quad\quad\quad\ \ 1 &\text{for}\quad n\geq3\
 \end{cases}\,,
 \label{eq:dnL/dBnratios}
\end{equation}
which highlights that the leading low-temperature contributions with zero and just one derivative for $\sqrt{2{\cal F}}$ exhibit the minimum suppression relatively to the corresponding zero-temperature ones in the strong magnetic/electric field limit.

\begin{figure}
 \centering
 \includegraphics[width=0.7\textwidth]{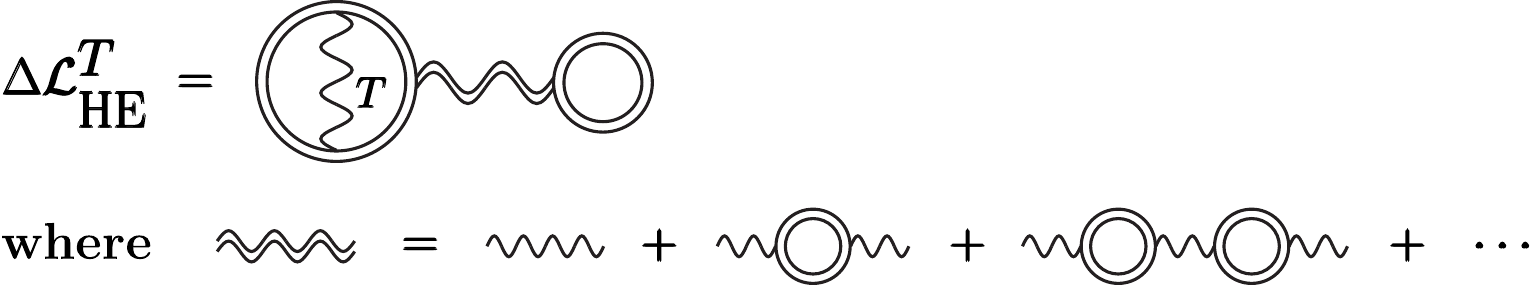}\vspace*{-1mm}
 \caption{Contribution to the Heisenberg-Euler Lagrangian encoding the leading strong magnetic/electric field behavior in the 1PR sector scaling quartically with temperature $T\ll m$.
 Here, the wiggly line with (without) the label $T$ denotes the purely thermal (zero-temperature) part of the tree-level photon propagator~\eqref{eq:D}.
 For the definition of the double solid line see Fig.~\ref{fig:1}.}
 \label{fig:2}
\end{figure}

Equation~\eqref{eq:dnL/dBnratios} in particular implies that the 1PR diagrams encoding (and effectively resumming) the leading strong magnetic/electric field behavior of the finite-temperature correction $\sim T^4$ for $T\ll m$ to the Heisenberg-Euler Lagrangian at any fixed loop order $\ell>2$ follow from its zero-temperature bubble-chain analogue \cite{Karbstein:2019wmj} by replacing ${\cal L}_{\rm HE}^{1\text{-loop}}\to{\cal L}_{\rm HE}^{2\text{-loop},T}$ in one of its two end loops that are characterized by a single derivative for $\sqrt{2{\cal F}}$. The possibility of choosing any of the two end loops should moreover result in an additional overall factor of two. It is clear from \Eqref{eq:dnL/dBnratios} that replacing ${\cal L}_{\rm HE}^{1\text{-loop}}\to{\cal L}_{\rm HE}^{2\text{-loop},T}$ in internal loops comes with a stronger suppression in the scaling with $\sqrt{2{\cal F}}$ and thus results in subleading contributions.
Correspondingly, all the individual diagrams predicted to dominate the 1PR sector in the strong magnetic/electric field limit are contained in the composite Feynman diagram depicted in Fig.~\ref{fig:2}. This diagram clearly receives contributions $\sim\alpha^{\ell-1}$ from all loop orders $\ell>2$.
Starting from the exact expression for the perturbative loop expansion of Heisenberg-Euler Lagrangian in Eq.~(2.39) of \cite{Karbstein:2023}, we have explicitly verified that this contribution to the Heisenberg-Euler effective Lagrangian in a generic constant electromagnetic field can be cast in the form
\begin{multline}
 \Delta{\cal L}_{\rm HE}^T=4\int\frac{{\rm d}^4k}{(2\pi)^4}\int\frac{{\rm d}^4k'}{(2\pi)^4}\,(2\pi)^4\delta(k)\\
 \times k^\rho\frac{\partial{\cal L}_{\rm HE}^{2\text{-loop},T}}{\partial F^{\rho\mu}} \bigl[\bigl(D^{-1}+\Pi^{1\text{-loop}}\bigr)^{-1}\bigr]^{\mu\nu}(k,k')\, \frac{\partial{\cal L}_{\rm HE}^{1\text{-loop}}}{\partial F^{\nu\sigma}} k'^\sigma\,, \label{eq:DeltaL}
\end{multline}
with tree-level photon propagator at zero temperature $D^{\mu\nu}$ and the one-loop photon-polarization tensor $\Pi^{1\text{-loop}}_{\mu\nu}$ given in \Eqref{eq:Pi}.

A comparison of \Eqref{eq:DeltaL} with its zero-field analogue given in Eq.~(2.159) of \cite{Karbstein:2023} confirms that the former indeed follows from the latter by replacing one of the derivatives of ${\cal L}_{\rm HE}^{1\text{-loop}}$ by a derivative of ${\cal L}_{\rm HE}^{2\text{-loop},T}$ and multiplying by $2$.
This implies that no new calculations are needed and the expression of \Eqref{eq:DeltaL} in a purely magnetic/electric field can be inferred by using these replacement rules.
Thereby, we find that the leading contribution to $\Delta{\cal L}_{\rm HE}^T$ in the strong magnetic/electric field limit at $\ell>2$ loops is given by
\begin{equation}
 \Delta{\cal L}_{\rm HE}^{\ell\text{-loop},T}=\frac{{\rm sgn}({\cal F})\pi}{540}\,T^4\alpha\,\frac{e\sqrt{2{\cal F}}}{m^2}\,\Bigl(\alpha\beta_1\ln\bigl(\tfrac{e\sqrt{2{\cal F}}}{m^2}\bigr)\Bigr)^{\ell-2}\,\Bigl[1+{\cal O}\Bigl(\ln^{-1}\bigl(\tfrac{e\!\sqrt{2{\cal F}}}{m^2}\bigr)\Bigr)\Bigr]+{\cal O}(T^6)\,, \label{eq:DeltaLlloop}
\end{equation}
where $\beta_1=1/(3\pi)$ is the coefficient of the QED beta function governing the running of the fine structure constant at one loop, and $\sqrt{2{\cal F}}$ either equals $|\vec{B}|$ or $-{\rm i}|\vec{E}|$.
The overall sign arises because the ratio ${\cal U}/{\cal F}$ changes sign when switching from a purely magnetic to a purely electric one.
Resumming the leading contributions in \Eqref{eq:DeltaLlloop} as $\Delta{\cal L}_{\rm HE}^T=\sum_{\ell=3}^\infty\Delta{\cal L}_{\rm HE}^{\ell\text{-loop},T}$, we arrive at the following all-order expression
\begin{equation}
 \Delta{\cal L}_{\rm HE}^T =\frac{{\rm sgn}({\cal F})\pi}{540}\,T^4\alpha\,\frac{e\sqrt{2{\cal F}}}{m^2}\,\alpha^{1\text{-loop}}\bigl(\tfrac{e\sqrt{2{\cal F}}}{m^2}\bigr)\,\beta_1\ln\bigl(\tfrac{e\sqrt{2{\cal F}}}{m^2}\bigr)\,\Bigl[1+{\cal O}\Bigl(\ln^{-1}\bigl(\tfrac{e\!\sqrt{2{\cal F}}}{m^2}\bigr)\Bigr)\Bigr]+{\cal O}(T^6)\label{eq:DeltaLlloopresum}
\end{equation}
for the strong magnetic/electric field limit of \Eqref{eq:DeltaL}, with the one-loop running of the fine structure given by
\begin{equation}
 \alpha^{1\text{-loop}}(\mu^2)=\frac{\alpha}{1-\alpha\beta_1\ln\bigl(\frac{\mu^2}{m^2}\bigr)}\,. \label{eq:alpha1loop}
\end{equation}
In passing, we note that the all-order expression given in \Eqref{eq:DeltaLlloopresum} is precisely of the form of $\Delta{\cal L}_{\rm HE}^{3\text{-loop},T}\sim\alpha^2$ in \Eqref{eq:DeltaLlloop}:
it can effectively be obtained from the latter by replacing one factor of $\alpha$ by $\alpha^{1\text{-loop}}\bigl(\tfrac{e\sqrt{2{\cal F}}}{m^2}\bigr)$.
This is in line with a phenomenological effective field theory (EFT) viewpoint suggesting that the couplings in a given diagram are to be evaluated at the relevant energy scales: for the coupling $\sim\alpha$ to the finite temperature $T\ll m$ this is the electron mass $m$, while for the effective coupling to the strong electromagnetic field (via the tadpole structure) this rather is $e\sqrt{2{\cal F}}$.

\section{Conclusions}\label{sec:concls}

In this note, we revisited the leading low-temperature $T\ll m$ correction to the Heisenberg-Euler Lagrangian in a perturbative loop expansion encoded in ${\cal L}_{\rm HE}^{2\text{-loops},T}$.
Resorting to a different formalism than the original work \cite{Gies:1999vb} where this correction was first studied, we showed that particularly the determination of the contribution $\sim T^4$ becomes essentially trivial.
Because it only requires taking derivatives of the one-loop Heisenberg-Euler Lagrangian ${\cal L}_{\rm HE}^{1\text{-loop}}$ at zero temperature it is straightforward to analytically extract the corresponding low-temperature correction $\sim T^4$ in any limit for which ${\cal L}_{\rm HE}^{1\text{-loop}}$ can be evaluated explicitly. This in particular also resulted in an exact expression for the imaginary part of ${\cal L}_{\rm HE}^{2\text{-loops},T}$ at quartic order in $T$ recovering the purely electric field case studied in \cite{Gies:1999vb} in the corresponding limit. Special attention is put on ${\cal L}_{\rm HE}^{2\text{-loops},T}$ at ${\cal O}(T^4)$ in the specific strong field limit characterized by ${\cal G}=0$ and $|{\cal F}|\gg(m^2/e)^2$ where -- in agreement with \cite{Gies:1999vb}  -- a scaling $\sim T^4\sqrt{2{\cal F}}$ is found. This is to be contrasted with the scaling $\sim{\cal F}\ln(e\sqrt{2{\cal F}})$ of both ${\cal L}_{\rm HE}^{1\text{-loop}}$ and ${\cal L}_{\rm HE}^{2\text{-loop}}$ in the zero-temperature limit.

In the next step, we then effectively dressed the two-loop 1PI contribution ${\cal L}_{\rm HE}^{2\text{loop},T}$ to the Heisenberg-Euler Lagrangian at ${\cal O}(T^4)$ by 1PR tadpole structures. This generates a subset of higher-loop 1PR contributions to  ${\cal L}_{\rm HE}^T$ in the low temperature limit $\sim T^4$. We extracted their leading behavior in the strong magnetic/electric field limit at any given loop order, and finally resum these to all orders.
While we are not aware of any indication that the 1PR sector studied here should dominate over its 1PI counterpart in the strong field limit, we believe that this is of interest in its own right.
In particular, this allowed us to explicitly demonstrate that the finite-temperature correction $\sim T^4$ to the Heisenberg-Euler Lagrangian generically receives contributions $\sim\alpha^{\ell-1}$ from all loop orders $\ell\geq2$.

Finally, we note that the very same considerations, of course, can be easily repeated for scalar QED yielding analogous finite-temperature corrections.
For recent progress in finite-temperature calculations \cite{Lowdon:2024atn} in quantum chromodynamics (QCD) with background electromagnetic fields see also \cite{Endrodi:2022wym,Endrodi:2026kmb}.

\acknowledgments

This work has been funded also by the Deutsche Forschungsgemeinschaft (DFG) under Grant No. 416607684 within the Research Unit FOR2783/2.
I would like to thank the organizers of the Hungarian-German WE-Heraeus-Seminar ``Particles and Plasmas in Strong Fields'' for inviting me to this both inspiring and enjoyable event at HZDR Dresden and in G\"orlitz.
My special thanks go to David Blaschke for convincing me to come up with a contribution to this special issue of {\it Particles}.

\end{document}